# Metamaterial superconductors

Igor I. Smolyaninov [1]) and Vera N. Smolyaninova [2])

*[1]Department of Electrical and Computer Engineering, University of Maryland, College Park, MD 20742, USA*

*[2]Department of Physics Astronomy and Geosciences, Towson University,*

*8000 York Rd., Towson, MD 21252, USA*

**Epsilon near zero (ENZ) conditions have been observed to enhance superconducting properties of a composite metamaterial based on random mixing of superconductor and ferroelectic nanoparticles. Here we analyse several other promising experimental geometries, which may considerably enhance electron pairing interaction in a metamaterial superconductor. The proposed geometries may be fabricated at the current level of nanotechnology development. They enable tuning of both frequency and spatial dispersion of the effective dielectric response function $\varepsilon_{eff}(q,\omega)$ of the metamaterial, thus enabling optimization of the metamaterial superconductor properties.**

## I. Introduction

Very recently it has been demonstrated that there exists a deep non-trivial connection between the fields of electromagnetic metamaterials and superconductivity research. Namely, we have demonstrated that the metamaterial approach to dielectric response engineering may increase the critical temperature of a composite superconductor-dielectric metamaterial [1,2]. Superconducting properties of a material, such as electron-electron interactions and the critical temperature of superconducting transition may be expressed via the effective dielectric response function $\varepsilon_{eff}(q,\omega)$ of the material [3]. Such a macroscopic electrodynamics description is valid if the material may be considered as a homogeneous medium on the spatial scales below the superconducting coherence length (the size of the Cooper pair), which equals ~100 nm in a typical BCS superconductor. Searching for natural materials exhibiting larger electron-electron interactions constitutes a traditional approach to high temperature superconductivity research. Very recently we pointed out that the newly developed field of electromagnetic metamaterials deals with somewhat related task of dielectric response engineering on sub-100 nm scale. Moreover, considerable enhancement of attractive electron-electron interaction may be expected in such actively studied metamaterial scenarios as epsilon near zero (ENZ) [4] and hyperbolic metamaterials [5]. In both cases $\varepsilon_{eff}(q,\omega)$ may become small and negative in substantial portions or the relevant four-momentum $(q,\omega)$ space, leading to enhancement of electron pairing interaction. This approach has been verified in experiments with compressed mixtures of tin and barium titanate nanoparticles of varying composition [2]. An increase of the critical temperature of the order of $\Delta T_c \sim 0.15$ K compared to bulk tin has been observed for 40% volume fraction of barium titanate nanoparticles, which corresponds to ENZ conditions. Similar results were also obtained with compressed mixtures of tin and strontium titanate nanoparticles. These highly non-trivial results clearly demonstrate a deep connection between superconductors and electromagnetic metamaterials.

Electromagnetic properties are known to play a very important role in the pairing mechanism of superconductors. According to the BCS theory [6], an electron Cooper pair is formed by two electrons which are loosely bound and have opposite spins. This process may be interpreted as attractive interaction through ionic lattice polarization, which these electrons create as they move through the lattice. Based on this interpretation, in the 1970s Kirzhnits *et al.* formulated their description of superconductivity in terms of the dielectric response function of the superconductor [3]. They demonstrated that electron-electron interaction in a superconductor may be expressed in the form of effective Coulomb potential

$$V(\vec{q},\omega) = \frac{4\pi e^2}{q^2 \varepsilon_{eff}(\vec{q},\omega)},$$  (1)

where $V=4\pi e^2/q^2$ is the Fourier-transformed Coulomb potential in vacuum, and $\varepsilon_{eff}(q,\omega)$ is the linear dielectric response function of the superconductor treated as an effective medium. Based on this approach they have derived expressions for the superconducting gap $\Delta$, critical temperature $T_c$, and other important parameters of the superconductor. Following the discovery of high $T_c$ superconductors in the next decade, detailed investigation of their electromagnetic properties confirmed that they indeed play a very important role in their electron pairing mechanism and charge dynamics [7].

In a parallel development, recent progress in plasmonics [8] and electromagnetic metamaterials [9] made us realize that an artificial medium or "metamaterial" may be engineered at will so that its effective dielectric response function $\varepsilon_{eff}(q,\omega)$ may conform to almost any desired behaviour. It appears natural to use this newly found freedom to engineer and maximize the electron pairing interaction (1) in an artificial "metamaterial superconductor" via engineering its dielectric response function $\varepsilon_{eff}(q,\omega)$. Since the superconducting coherence length (the size of the Cooper pair) is $\xi$~100 nm in a typical BCS superconductor, we have an opportunity to engineer the fundamental metamaterial building blocks in such a way that the effective electron-electron interaction (1) will be maximized, while homogeneous treatment of $\varepsilon_{eff}(q,\omega)$ will remain valid. In order to do this, the metamaterial unit size must fall within a rather large window between ~0.3 nm (given by the atomic scale) and $\xi \sim 100$ nm scale of a typical Cooper pair. However, because of such a fine scale, this task is much more challenging compared to typical applications of superconducting metamaterials [10]. A metamaterial superconductor geometry must be chosen carefully in such a way, that its fabrication requirements may be met at the current level of nanotechnology development. The two solutions suggested so far [1,2] involve either a random mixture of superconductor (tin) and ferroelectric (BaTiO$_3$) nanoparticles mixed in such a proportion that the random mixture exhibits negative ENZ behaviour, or the hyperbolic metamaterial scenario, which requires layer-by-layer fabrication of a planar multilayer superconductor-ferroelectric metamaterial. Both scenarios naturally lead to broadband small and negative $\varepsilon_{eff}(q,\omega)$ behavior in substantial portions of the relevant four-momentum spectrum. In the isotropic ENZ scenario the effective pairing interaction is described by Eq. (1), where $\varepsilon_{eff}(q,\omega)$ of the nanoparticle mixture may be calculated based on the Maxwell-Garnett approximation [11]. On the other hand, in the hyperbolic metamaterial scenario the effective Coulomb potential from Eq. (1) assumes the form

$$V(\vec{q},\omega) = \frac{4\pi e^2}{q_z^2 \varepsilon_2(\vec{q},\omega) + \left(q_x^2 + q_y^2\right)\varepsilon_1(\vec{q},\omega)} \quad , \tag{2}$$

where $\varepsilon_{xx} = \varepsilon_{yy} = \varepsilon_1$ and $\varepsilon_{zz} = \varepsilon_2$ have opposite signs [1]. As a result, the effective Coulomb interaction of two electrons may become attractive and very strong along spatial directions where

$$q_z^2 \varepsilon_2(\vec{q},\omega) + \left(q_x^2 + q_y^2\right)\varepsilon_1(\vec{q},\omega) \approx 0 \tag{3}$$

and negative, which indicates that the superconducting order parameter must be strongly anisotropic. This situation resembles the hyperbolic character of such high $T_c$ superconductors as BSCCO [12]. Experimental realization of the ENZ scenario [2] indeed demonstrated an increase of the critical temperature of the tin-based metamaterial superconductor compared to the bulk tin, and provided a proof of principle of the metamaterial superconductor approach. In this paper we consider two alternative metamaterial superconductor geometries, which may provide us with additional powerful tools of dielectric response engineering.

## II. ENZ metamaterials based on random nanoparticle mixtures

Let us start with a brief overview of electromagnetic and superconducting properties of ENZ metamaterials fabricated as random nanoparticle mixtures [2]. According to the Maxwell-Garnett approximation [11], random mixing of nanoparticles of a superconducting "matrix" with dielectric "inclusions" (described by the dielectric constants $\varepsilon_m$ and $\varepsilon_i$, respectively) results in the effective medium with a dielectric constant $\varepsilon_{eff}$, which may be obtained as

$$\left( \frac{\varepsilon_{eff} - \varepsilon_m}{\varepsilon_{eff} + 2\varepsilon_m} \right) = \delta_i \left( \frac{\varepsilon_i - \varepsilon_m}{\varepsilon_i + 2\varepsilon_m} \right), \tag{4}$$

where $\delta_i$ is the volume fraction of the inclusions (assumed to be small). The explicit expression for $\varepsilon_{eff}$ may be written as

$$\varepsilon_{eff} = \varepsilon_m \frac{(2\varepsilon_m + \varepsilon_i) - 2\delta_i(\varepsilon_m - \varepsilon_i)}{(2\varepsilon_m + \varepsilon_i) + \delta_i(\varepsilon_m - \varepsilon_i)} \tag{5}$$

The ENZ condition ($\varepsilon_{eff} \approx 0$) are obtained around

$$\delta_i = \frac{2\varepsilon_m + \varepsilon_i}{2(\varepsilon_m - \varepsilon_i)}, \tag{6}$$

which means that $\varepsilon_m$ and $\varepsilon_i$ must have opposite signs, and $\varepsilon_i \approx -2\varepsilon_m$ so that $\delta_i$ will be small. Eq.(6) may also be written as

$$\varepsilon_m = -\varepsilon_i \frac{1 + 2\delta_i}{2(1 - \delta_i)} \tag{7}$$

Dielectric permittivity $\varepsilon_{m=}\varepsilon_m(0, \omega)$ of the metal component is typically given by the Drude model in the far infrared and THz ranges as

$$\varepsilon_m = \varepsilon_{m\infty} - \frac{\omega_p^2}{\omega(\omega + i\gamma)} \approx -\frac{\omega_p^2}{\omega(\omega + i\gamma)}, \tag{8}$$

where $\varepsilon_{m\infty}$ is the dielectric permittivity of metal above the plasma edge, $\omega_p$ is its plasma frequency, and $\gamma$ is the inverse free propagation time. Since $\varepsilon_m$ is large and negative, the dielectric permittivity $\varepsilon_i$ of inclusions must be positive and very large. Ferroelectric materials, such as BaTiO$_3$, which have large positive $\varepsilon_i$ in the THz range appear to be a very good choice of such a dielectric. Moreover, if the high frequency behavior of $\varepsilon_i$ may be assumed to follow the Debye model [13]:

$$\mathrm{Re}\,\varepsilon_i = \frac{\varepsilon_i(0)}{1 + \omega^2 \tau^2} \approx \frac{\varepsilon_i(0)}{\omega^2 \tau^2}, \tag{9}$$

a broadband ENZ behaviour results due to the similar $\sim\omega^{-2}$ functional behaviour of $\varepsilon_i$ and $\varepsilon_m$ in the THz range (compare Eqs. (8) and (9)).

At the microscopic level Eq.(7) may be interpreted as metamaterial-induced tuning of spectra of various bosonic excitations (acoustic phonons, optical phonons, plasmons, etc.) which may participate in the electron pairing interaction. Following [14], a simplified dielectric response function of a metal may be written as

$$\varepsilon_m(q, \omega) = \left(1 - \frac{\omega_p^2}{\omega^2 - \omega_p^2 q^2 / k^2}\right)\left(1 - \frac{\Omega_1(q)}{\omega^2}\right)..\left(1 - \frac{\Omega_n^2(q)}{\omega^2}\right) \tag{10}$$

where $\omega_p$ is the plasma frequency, $k$ is the inverse Thomas-Fermi radius, and $\Omega_n(q)$ are dispersion laws of various phonon modes. Zeroes of the dielectric response function of the bulk metal (which correspond to various bosonic modes) maximize electron-electron interaction given by eq.(1). Compared to the bulk metal, zeroes of the effective dielectric response function $\varepsilon_{eff}(q,\omega)$ of the metal-dielectric metamaterial are observed at shifted positions given by Eq.(7) as

$$\left(1-\frac{\omega_p^2}{\omega^2-\omega_p^2q^2/k^2}\right)\left(1-\frac{\Omega_1(q)}{\omega^2}\right)...\left(1-\frac{\Omega_n^2(q)}{\omega^2}\right)=-\varepsilon_i\frac{1+2\delta_i}{2(1-\delta_i)} \qquad (11)$$

The latter equation is easy to analyse qualitatively if the bosonic modes are far away from each other and $\varepsilon_i$=const. If $\omega>>\Omega_n$ the dispersion law of the metamaterial plasmon may be obtained as

$$\omega^2=\frac{\omega_p^2}{\left(1+\varepsilon_i\frac{1+2\delta_i}{2(1-\delta_i)}\right)}+\omega_p^2\frac{q^2}{k^2} \quad , \qquad (12)$$

which tend to the plasmon dispersion law of bulk metal in the limit $\varepsilon_i\rightarrow0$. Thus, as expected, presence of the dielectric component in the metamaterial lowers its effective plasma frequency. In the case of ferroelectrics, which low frequency dielectric constant is very large ($\varepsilon_i\sim10^3$-$10^4$), the effective plasma frequency of the metamaterial becomes comparable with the Debye frequency of phonons in metal. As a result, plasmonic mechanism of superconductivity may be realized in a metamaterial superconductor, while BCS-like phonon mediated mechanism appears to be suppressed. Indeed, if for the sake of simplicity we consider a metal which supports only one phonon mode $\Omega(q)$, and assume that $\omega<<\omega_p$, Eq.(11) may be simplified to produce an expression for the effective phonon mode of the metamaterial:

$$\omega^2=\frac{\Omega^2(q)}{\left(1+\frac{\varepsilon_iq^2(1+2\delta_i)}{2(q^2+k^2)(1-\delta_i)}\right)} \qquad (13)$$

Similar to plasmons, presence of the dielectric component in the metamaterial lowers its phonon frequencies. According to BCS theory, lower phonon modes, and therefore lower effective Debye frequency should lead to lower $T_c$. The fact that the critical temperature is increased in experiments with tin/$BaTiO_3$ metamaterial superconductors [2] points toward some role of plasmon-mediated electron pairing in these metamaterials. Another important implication of Eq.(13) is that metamaterial approach enables tuning of both plasmon and phonon modes of the material.

### III. Core-shell ENZ metamaterial superconductors

While experiments with random nanoparticle mixtures [2] provided a conclusive proof of critical temperature increase in an ENZ metamaterial, the random nanoparticle mixture geometry is not ideal. It is clear that simple mixing of superconductor and dielectric nanoparticles leaves room for substantial spatial variations of $\delta_t$ throughout a macroscopic metamaterial sample, leading to considerable broadening of the superconducting transition. We suggest to resolve this problem by implementing the plasmonic core-shell metamaterial geometry, which was recently proposed to achieve partial cloaking of macroscopic objects [15]. The cloaking effect relies on mutual cancellation of scattering by the dielectric core and plasmonic shell of the nanoparticle, so that the effective dielectric constant of the nanoparticle becomes very small and close to that of vacuum. This approach may be naturally extended to core-shell nanoparticles having negative ENZ behaviour. Synthesis of such individual ENZ core-shell nanostructures based, for example, on $BaTiO_3$ cores and tin shells (shown in Fig. 1(a)), followed by nanoparticle self-assembly into a bulk ENZ metamaterial appears to be a viable way to fabricate a metamaterial superconductor.

The design of an individual core-shell nanoparticle is based on the fact that scattering of electromagnetic field by a subwavelegth object is dominated by its electric dipolar contribution, which is defined by the integral sum of its volume polarization [15]. A material with $\varepsilon > 1$ has a positive electric polarizability, while a material with $\varepsilon < 1$ has a negative electric polarizability (since the local electric polarization vector $P = (\varepsilon - 1)E/4\pi$ is opposite to $E$). As a result, the presence of a plasmonic shell cancels the scattering produced by the core, thus providing a cloaking effect. Similar consideration for the ENZ case leads to the following condition for the core-shell geometry:

$$r_1^3 \varepsilon_i = -\left(r_2^3 - r_1^3\right)\varepsilon_m ,  \qquad (14)$$

where $r_1$ and $r_2$ are the radii of the dielectric core and the metal shell, respectively. Our numerical calculations of THz wave scattering performed for an example of tin-BaTiO$_3$ core-shell particle satisfying Eq.(14) confirm its ENZ character. The results of our scattered power calculations are shown in Figs.1(b-e). They were performed at $\nu$=1THz based on the measured dielectric properties of Sn [16] and BaTiO$_3$ [17] extrapolated by Eqs. (8) and (9), respectively. Because of numerical mesh limitations, our calculations were performed for a micrometer scale core-shell particle satisfying Eq.(14). However, our results are scalable to a nanometer scale core-shell geometry. In our numerical simulations shown in Fig.1 we have compared THz scattering from a core-shell particle in Fig.1(e) with a series of field plots in Figs.1(b-d) obtained for homogeneous particles with a given constant dielectric permittivity $\varepsilon$. Our result for a core-shell particle satisfying Eq.(14) indeed looks very similar to the result for a homogeneous particle having $\varepsilon = $ - 0.01, which is shown in Fig.1(c).

We should also point out that in addition to obvious advantage in homogeneity, a core-shell based metamaterial superconductor design enables tuning of spatial dispersion of the effective dielectric permittivity $\varepsilon_{eff}(q,\omega)$ of the metamaterial, which would further optimize its $T_c$. Indeed, since spatial dispersion of dielectrics is typically very weak, Eq.(7) indicates that the ENZ metamaterial design would benefit from compensation of spatial dispersion of the dielectric response function of metal $\varepsilon_m(q,\omega)$. Spatial dispersion of a metamaterial is indeed well known to originate from plasmonic effects in its metallic constituents. In a periodic core-shell nanoparticle-based ENZ metamaterial spatial dispersion is defined by coupling of plasmonic modes of its individual nanoparticles. This coupling enables propagating plasmonic Bloch modes and, hence, nonlocal effects. Examples of such spatial dispersion calculations in nanoparticle chains may be found in [18]. The dispersion law of such Bloch modes is typically obtained as

$$\omega^2 = \omega_0^2 + 2\gamma_i \omega_1^2 \cos(qd), \qquad (15)$$

where $\omega_0$ is the plasmon eigenmode frequency of the individual nanoparticle, $\omega_1$ is the nearest neighbour coupling, $d$ is the inter-particle distance, and $\gamma_i$ is the polarization-dependent constant ($\gamma_T=1$ and $\gamma_L=-2$ for the transverse and longitudional modes, respectively) [19]. Thus, by tuning the core-shell nanoparticle radius $r_2$ we will tune the average inter-particle distance $d=2r_2$ in Eq.(15) and therefore, tune spatial dispersion of the metamaterial. According to Eqs.(12,13), in the absence of spatial periodicity both plasmon and phonon modes of the composite metal-dielectric metamaterial exhibit considerable spatial dispersion (see Fig.2(b)). As evident from Eq.(15), Bragg scattering of the plasmons and phonons by the periodic core-shell metamaterial structure provides an obvious way to compensate spatial dispersion of these modes leading to broadband ENZ behaviour. We have confirmed these qualitative arguments by numerical calculations of plasmon dispersion law of the tin/BaTiO$_3$ ENZ metamaterial using Eq.(12), and its modification in a $d$=50 nm core-shell design. Material parameters of tin and BaTiO$_3$ [16,17], as well as experimentally measured value of $\delta_i$=0.4 [2] have been used in these calculations presented in Fig.2(b). Our numerical results indicate that spatial dispersion of $\varepsilon_{eff}(q,\omega)$ may indeed be suppressed in the core-shell metamaterial design.

## IV. Metamaterial superconductors based on artificial high $\varepsilon$ metamaterials

While random ENZ nanoparticle mixtures [2] and the core-shell geometries described above appear to produce promising metamaterial superconductor designs, both approaches rely on natural high dielectric permittivity materials, such as BaTiO$_3$. Unfortunately, the dielectric constant $\varepsilon_i$ of all natural dielectrics becomes rather modest at higher frequencies, so that the metal-dielectric ENZ designs relying on $\varepsilon_i \sim -\varepsilon_m$ cannot be realized using natural dielectrics. Fortunately, this difficulty may be resolved by implementing artificial high $\varepsilon$ metamaterials, which have been developed in a number of recent theoretical proposals and experiments [20,21].

Historically, the suggestion to use highly polarizable dielectric side chains or layers in order to increase T$_c$ of a 1D or 2D electronic system can be traced back to pioneering papers by W. Little [22] and V. Ginzburg [23], which predicted possible existence of room temperature superconductivity in such systems. While these proposals never led to hypothesized room temperature superconductivity, experimental attempts to realize such superconductors were limited by modest polarizability of natural materials. Recent development of high $\varepsilon$ metamaterials may give these proposals another chance. Unlike the phonon-mediated electron pairing in the BCS theory, alternative mechanisms of superconductivity in these geometries rely on either exciton or plasmon-mediated pairing interaction. As summarized in [24], the critical temperature of the BCS superconductor is typically derived as

$$T_c = \theta_D \exp(-\frac{1}{\lambda_{eff}}) \,, \qquad (16)$$

where $\theta_D$ is the Debye temperature and $\lambda_{eff} = N(0)V$, where $N(0)$ is the electronic level density near the Fermi surface in the normal state and $V$ is some average matrix element of electron interaction which corresponds to attraction. If the phonon mechanism of attraction between conduction electrons is replaced with some other mechanism, such as excitonic or plasmonic, $\theta_D$ in equation (16) need to be replaced with much higher $\theta_{ex} = \hbar\omega_{ex}/k$ or $\theta_{pl} = \hbar\omega_{pl}/k$ characteristic temperatures, which may lead to considerable increase of Tc.

As pointed out in e.g. [25], plasmon-mediated pairing of electrons may be understood in terms of image charge-mediated Coulomb interaction. This point of view allows us to illustrate the potential advantage of artificial high $\varepsilon$ metamaterial in a thin film (sandwich) geometry proposed by Ginzburg. Let us consider two electrons located next to the interface between two media with dielectric permittivities $\varepsilon_1$ and $\varepsilon_2$, as shown in Fig.3. The field acting on charge $e_2$ in medium $\varepsilon_1$ at $z>0$ is obtained as a superposition of fields produced by charge $e_1$ and its image $e_1'$ [26]. As a result, the effective Coulomb potential may be obtained as

$$V = \frac{e}{\varepsilon_1}\left(\frac{e}{r_1} - \frac{e}{r_2}\left(\frac{\varepsilon_2 - \varepsilon_1}{\varepsilon_2 + \varepsilon_1}\right)\right) \,, \qquad (17)$$

which may be simplified as

$$V = \frac{2e^2}{r(\varepsilon_2 + \varepsilon_1)} \qquad (18)$$

if both charges are located very close to the interface, so that $r_1 = r_2 = r$. The $\varepsilon_2 = -\varepsilon_1$ condition, which maximizes pairing interaction, corresponds to the dispersion law of surface plasmons propagating along the interface. Thus, an artificial metamaterial dielectric with large enough positive dielectric constant $\varepsilon_i$ may enhance superconductivity in a thin metal film via surface plasmon-mediated electron pairing. Large enough $\varepsilon_i$ lowers the surface plasmon energy, so that plasmons may indeed play considerable role in the electron pairing mechanism.

An example of high $\varepsilon$ metamaterial geometry, which is amenable to lithographic fabrication using currently available 14 nm node technology [27] is presented in Fig.4(a). It is based on the approach to high $\varepsilon$ THz metamaterials, which was experimentally developed in [21]. In this work a peak dielectric constant of ~1000 at $\nu$~1THz with a lower frequency quasi-static value of over 400 were experimentally obtained by drastically increasing the effective permittivity through strong capacitive coupling within the metamaterial unit cell, and decreasing the diamagnetic response of the cell, while maintaining low losses. Since the metamaterial unit size must be kept below the superconducting coherence length $\xi$~100 nm, the geometry used in [21] has to be scaled down to the currently available 14 nm fabrication node. The unit cell structure of the high-index metamaterial shown in Fig.4(a) is made of a thin 'I'-shaped metallic patch symmetrically embedded in a dielectric material. The 3D structure of this metamaterial may be described as an array of subwavelength capacitors, in which the dimensions of "I", as well as the gap width (defined by $g = L-a$; see Fig. 4(a)) and the interlayer spacing $d$ define the capacitor values. The effective dielectric constant of the metamaterial may be estimated as

$$\varepsilon_{eff} \approx \varepsilon_s + \frac{4\pi P}{E}, \qquad (19)$$

where $\varepsilon_s$ is the dielectric constant of the substrate (polyimide substrate with $\varepsilon_s \sim 3.2$ was used in [21]), and the polarization density $P$ may be approximated as the dipole moment per unit volume as

$$P = \frac{Qa}{L^2 d} \qquad (20)$$

Under the influence of external driving field $E$, a large amount of surface charge $Q$ is accumulated on each arm of the metallic patch capacitor as the charges in each arm interact with opposite charges in close proximity across the small gap $g$. Charge accumulation on the edges of the metallic patch leads to an extremely large dipole moment in the unit cell, as the accumulated charge is inversely proportional to the gap width $g$ [21]:

$$Q \propto \frac{\varepsilon_s L^3}{4\pi g} E \qquad (21)$$

As a result, very large values of $\varepsilon_{eff}$

$$\varepsilon_{eff} \propto \varepsilon_s \left( 1 + \frac{aL}{dg} \right) \qquad (22)$$

may be obtained in the low frequency quasi-static approximation. Eq.(22) demonstrates that the high $\varepsilon$ metamaterial geometry implemented in [21] may indeed be scaled down to spatial scales below $\xi \sim 100$ nm without considerable loss in performance. While $L/g \sim 26$ ratio implemented in [21] ($L=40$ μm, $g=1.5$ μm) cannot be achieved using the 14 nm fabrication node under the $L<\xi$ constraint, a germanium substrate would have considerably larger value of $\varepsilon_s \sim 16$ compared to $\varepsilon_s \sim 3.2$ of polyimide substrate used in [21]. Assuming germanium substrate and $L/g \sim 7$ ratio feasible for the 14 nm fabrication node ($L=98$ nm, $g=14$ nm, $w=28$ nm in the geometry shown in Fig.4(a)), based on Eq.(22) we may realistically achieve $\varepsilon_{eff} \sim 150$ value in the quasi-static limit. Real and imaginary parts of $\varepsilon_{eff}$ numerically calculated for these geometrical parameters are shown in Fig.4(b). Broadband high $\varepsilon$ behaviour of the metamaterial has been confirmed in these calculations, while metamaterial losses remain modest below 50 THz. We anticipate that using such high $\varepsilon$ metamaterial in a sandwich geometry proposed by Ginzburg may considerably enhance $T_c$ in a thin superconductor layer deposited onto high $\varepsilon$ metamaterial.

## V. Conclusion

In conclusion, we have analyzed several promising experimental geometries, which may considerably enhance electron pairing interaction in a metamaterial superconductor. While the metamaterial unit size of the proposed designs is below the superconducting coherence length $\xi \sim 100$ nm of a typical BCS superconductor, their fabrication requirements may be met at the current level of nanotechnology development. The proposed geometries enable tuning of both frequency and spatial dispersion of the effective dielectric response function $\varepsilon_{eff}(q, \omega)$ of the metamaterial, thus enabling optimization of the metamaterial superconductor properties.

## Acknowledgement

This work was supported in part by NSF grant DMR-1104676.

**Figure Captions**

**Figure 1.** (a) ENZ core-shell nanoparticle based on BaTiO$_3$ core and tin shell. Such core-shell nanoparticles may be used in an alternative metamaterial superconductor geometry. (b-e) Numerical comparison of THz scattering from an ENZ core-shell particle (e) with a series of field plots in (b-d) obtained for homogeneous particles with a given constant dielectric permittivity $\varepsilon$. Spatial distribution of scattered power has been calculated at $\nu$=1THz. The dielectric constants $\varepsilon$ of homogeneous particles in (b-d) equal -2, -0.01, and +2, respectively. Our result (e) for an ENZ core-shell particle satisfying Eq.(14) indeed looks very similar to the result for a homogeneous particle having $\varepsilon$ = - 0.01 shown in (c).

**Figure 2**. (a) Schematic geometry of the ENZ metamaterial superconductor based on the core-shell nanoparticles shown in Fig.1(a). (b) Numerical calculations of plasmon dispersion law of the tin/BaTiO3 ENZ metamaterial using Eq.(12), and its modification in a $d$=50 nm core-shell design. Material parameters of tin and BaTiO3 [16,17], as well as experimentally measured value of $\delta\tilde{s}$=0.4 [2] have been used in these calculations.

**Figure 3**. Schematic view of the thin film (sandwich) geometry: an electron $e_2$ located next to the interface between two media with dielectric permittivities $\varepsilon_1$ and $\varepsilon_2$, interacts with electron $e_1$ and its image $e_1$'.

**Figure 4**. (a) Unit cell structure of the high-index metamaterial, made of a thin 'I'-shaped metallic patch symmetrically embedded in a dielectric material. The 3D structure of this metamaterial may be described as an array of subwavelength capacitors, in which the gap width defined by $g$=$L$-$a$ and the interlayer spacing $d$ (exaggerated for clarity) define the capacitor values. (b) Real and imaginary parts of $\varepsilon_{eff}$ numerically calculated for the metamaterial geometry shown in (a) for the following

values of geometrical parameters: $L$=98 nm, $g$=14 nm, $w$=28 nm. The metamaterial exhibits broadband high $\varepsilon$ behaviour, while metamaterial losses remain modest below 50 THz.

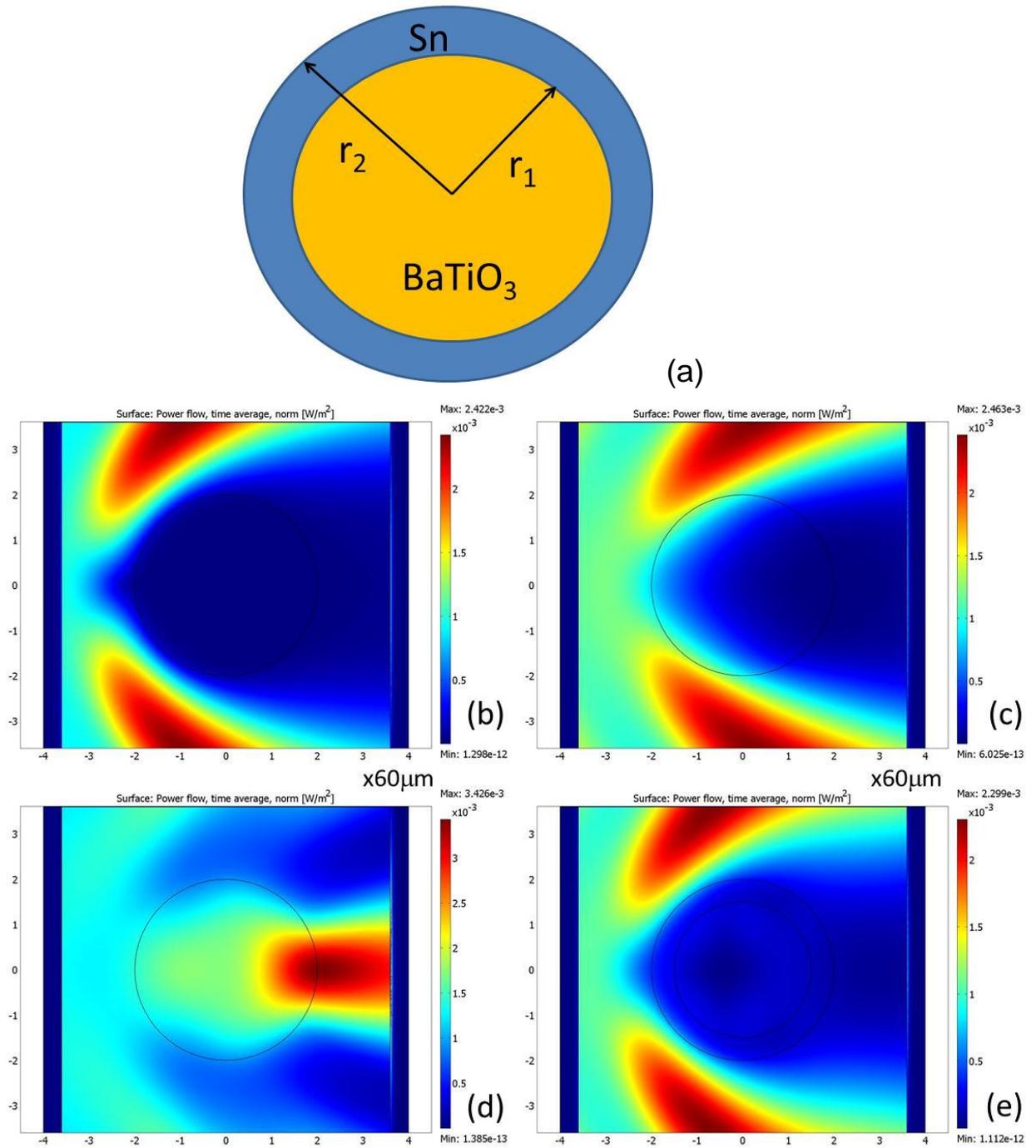

(a)

(b)

(c)

x60µm

(d)

(e)

x60µm

Fig.1

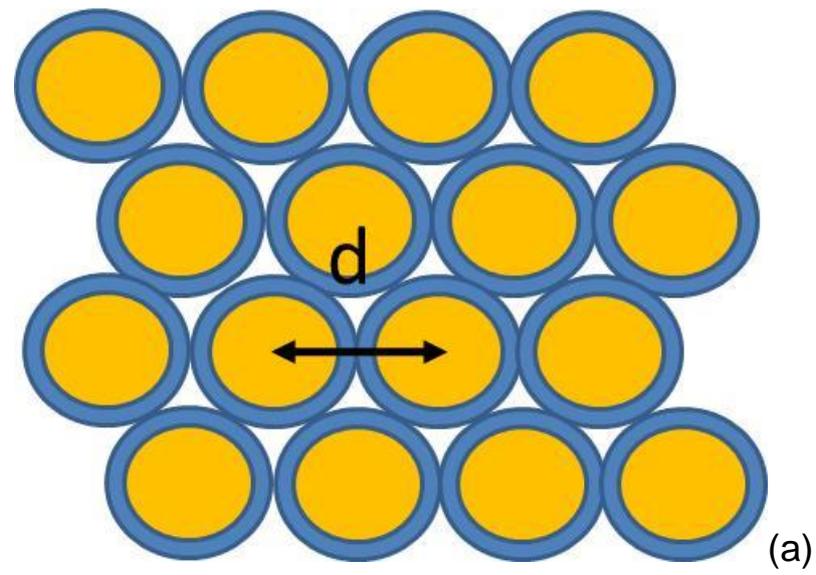

(a)

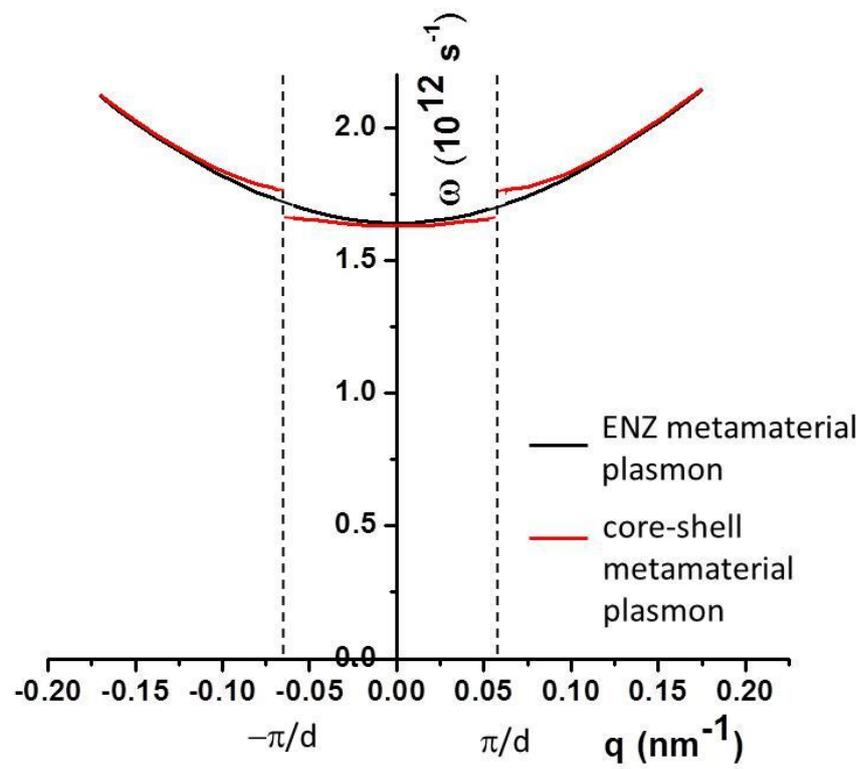

(b)

Fig.2

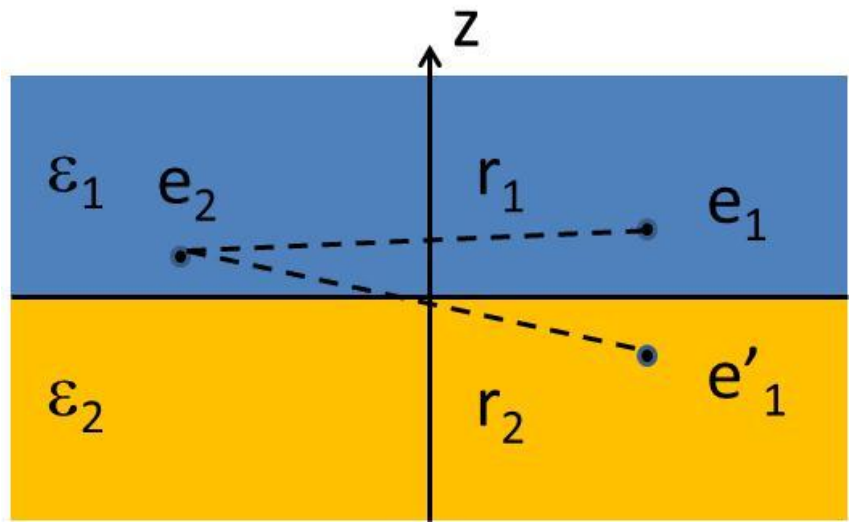

Fig. 3

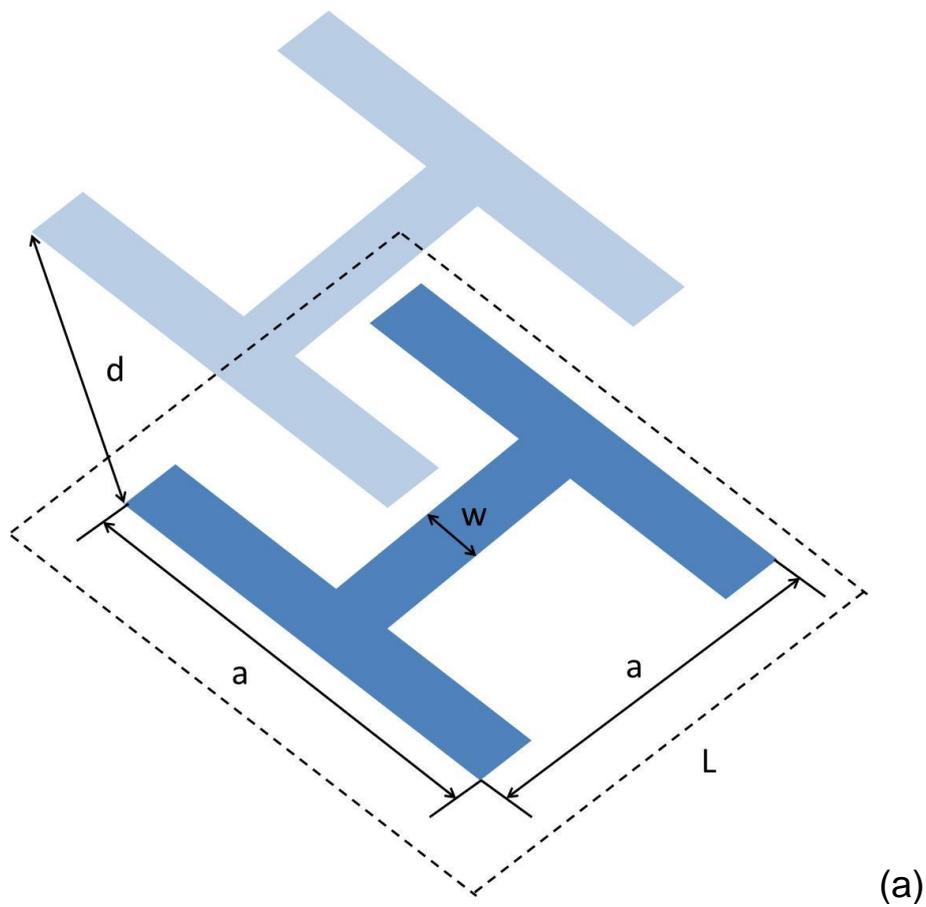

(a)

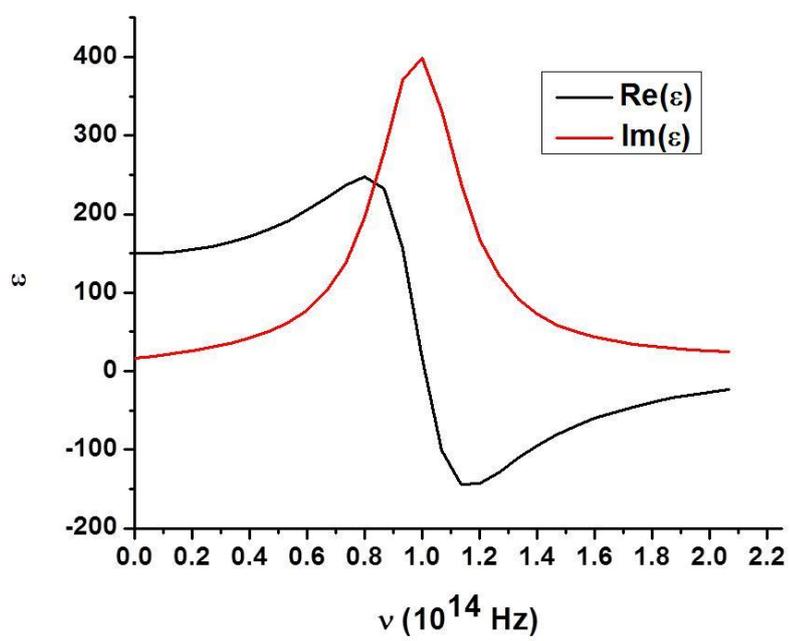

(b)

Fig. 4